\documentclass[]{iopart}
\usepackage{graphicx}
\usepackage{bm}
\usepackage{wasysym}
\begin{document}

\title[Disorder by disorder and quantum confinement in the quantum Ising model]{Disorder by disorder and confinement in the quantum Ising model in the pyrochlore lattice}
\author{Chyh-Hong Chern}
\eads{\mailto{chchern@ntu.edu.tw}}
\address{Department of Physics, Center for Theoretical Sciences, and Center for Quantum Science and Engineering, National Taiwan University, Taipei 10617, Taiwan}
\author{Chen-Nan Liao and Yang-Zhi Chou}
\address{Department of Physics, National Taiwan University, Taipei 10617, Taiwan}
\date{\today}
\begin{abstract}
At zero temperature, the classical antiferromagnetic Ising model on the pyrochlore lattice is a spin disorder phase of the critical spin correlation.  It is a deconfined phase in that the binding energy of the monopole-anti-monopole pair is independent of their distance of separation.  We show that turning on a transverse magnetic field turns it into the cooperative paramagnet, and the spin correlation becomes exponential decay.  Furthermore, it introduces the quantum confinement (of magnetic monopoles), where the binding energy of the pair is proportional to their distance of separation.  This disorder state undergoes adiabatic transition to the paramagnetic state in the large field limit.  The effective Hamiltonian (without magnetic monopoles) in small field is the Ising Hamiltonian plus ring exchange interaction.
 \end{abstract}

\maketitle

Quantum number fractionalization is one of the most intriguing topics in physics.  In particular, searching for the magnetic monopole has been of persistent interest for centuries.  This interest was recently aroused again by the discovery of the magnetic-monopole-like spin excitations in the spin ice systems~\cite{Morris2009science,Fennell2009nature}.  The energy of the monopole--anti-monopole pair was shown to be inversely proportional to their distance of separation, \emph{i.e.} the Coulomb phase of the magnetic monopole~\cite{Castelnovo2007nature}.

\begin{figure}[htb]
\includegraphics[width=0.4\textwidth]{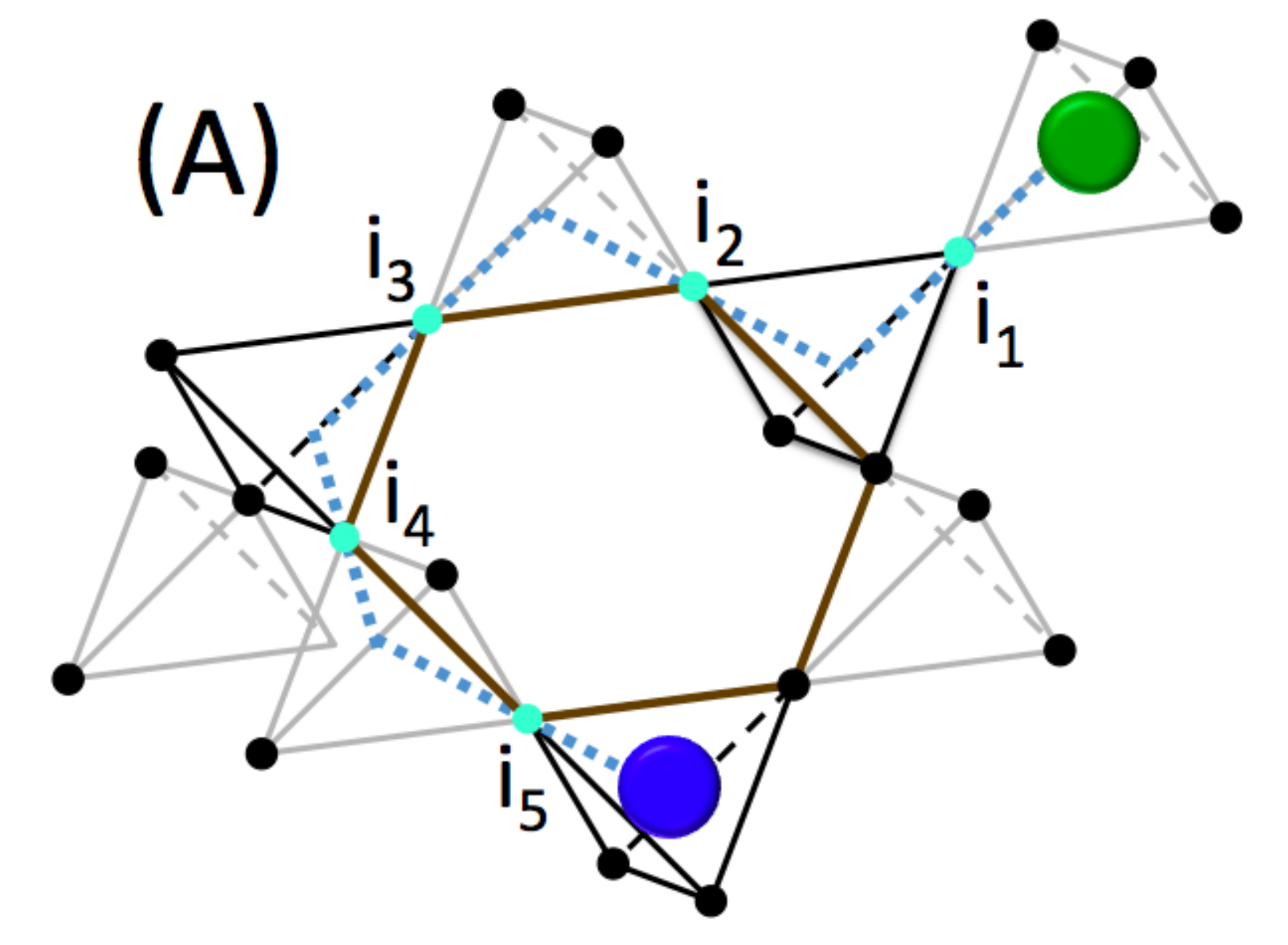}
\includegraphics[width=0.5\textwidth]{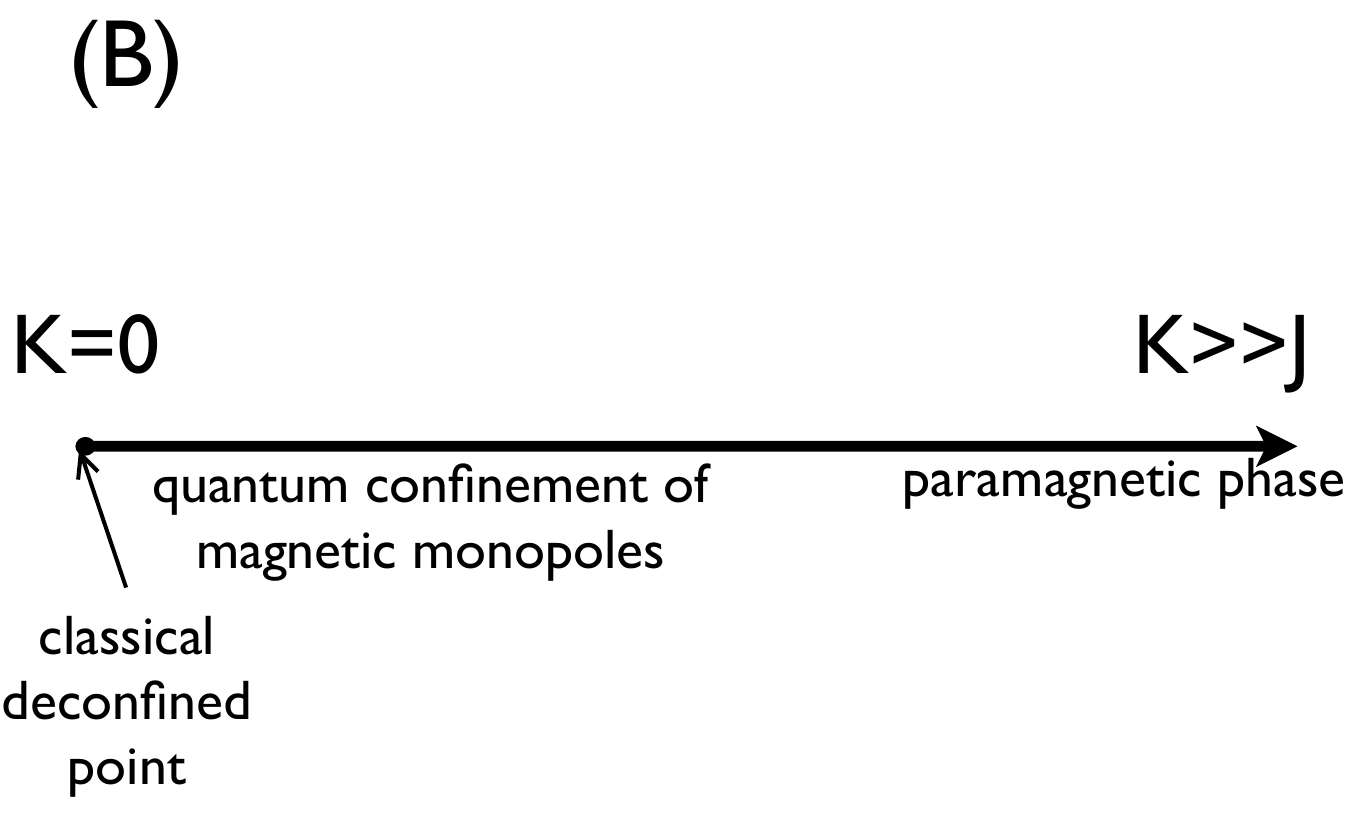}
\caption{(Color online) (a) A cartoon picture for the string excitation.  Flipping spin at site $i_1$ generates a monopole (green ball) and anti-monopole (blue ball) pair.  Further flipping at site $i_2$ makes the anti-monopole hop.  In this figure, the anti-monopole hops between 5 tetrahedra following the blue dash line, which is the flux-tube excitation of the energy $\sim l \, \frac{K^2}{J}$.  The line in brown is an example of the pyrochlore hexagon which is the shortest path by which monopole and the anti-monopole can annihilate.  (b) The phase diagram of the antiferromagnetic quantum Ising model in the pyrochlore lattice at zero longitudinal magnetic field.  Deconfinement only occurs at $K=0$.  With non-vanishing $K$, the system becomes the quantum confined phase and undergoes a continuous transition to the paramagnetic phase.}\label{Fig:string}
\end{figure}

The spin ice systems are found in the pyrochlore lattice.  It is a 3-dimensional system in which tetrahedra are stacked by sharing their corners as shown in Fig.~(\ref{Fig:string}a).   The ground state of the spin ice system is that the spin orientations in each tetrahedron are pointing "2-in--2-out" of the tetrahedron~\cite{Bramwell2001science,Castelnovo2007nature}, similar to the ice rule~\cite{Pauling1935}.  For a given classical spin configuration in the ground state manifold, one can create a spin excitation by flipping a spin.  It results in breaking the ice rule at two neighboring tetrahedra.  In one of them, the spin orientations will be ``1-in--3-out", giving rise to a monopole, and in the other they will be ``3-in--1-out", giving rise to an anti-monopole.  One can try to hop the anti-monopole by flipping another spin pointing inward to be outward in the anti-monopole tetrahedron.  If only the Ising interaction is taken into account, the distance of separation of the monopole and anti-monopole can increase with no cost in energy.  In other words, the pair is deconfined because the binding energy does not depend on their distance of separation.  When considering the classical dipolar interaction additionally, the binding energy becomes inversely proportional to their distance of separation~\cite{Castelnovo2007nature}.

Unlike ordinary electrons, the emergent magnetic monopole and anti-monopole in the spin ice systems are \emph{classical} particles~\cite{Castelnovo2007nature}.  It is important to investigate whether their \emph{quantum} counterparts are deconfined or not.  Searching for the \emph{quantum} deconfinement and quantum spin disorder phase, namely the spin liquid, has attracted tremendous attention in the last few decades~\cite{Hermele2004prb,Senthil2004science,sikora2009prl,fulde2008,nussinov2007prb,qi2009science}. 
The questions are, however, extremely difficult in the 3-dimensional system, because the quantum fluctuation is largely suppressed in three dimensions and, in addition, the limits on computational power also restricts the calculation from being conclusive.  In this Letter, we study the antiferromagnetic quantum Ising model
\begin{eqnarray}
H = J\sum_{\langle ij\rangle}S^z_iS^z_j -K\sum_i S^x_i \label{eq:hamiltonian}
\end{eqnarray}
where $\langle ij\rangle$ sums over all nearest neighbor spins, $i$ sums over all spins, $J>0$, and $S^k$ are the Pauli spin matrices for spin-1/2.  We will numerically demonstrate that the ground state of Eq.~(\ref{eq:hamiltonian}) is spin disorder with an exponential decay correlation, \emph{i.e.} disorder from disorder~\cite{Moessner2001prb,Moessner2000prl}.  The binding energy of the monopole--anti-monopole pair will be calculated by using the generalized random phase approximation.  In the small field limit, it is $\sim l \, \frac{K^2}{J}$, and $\sim l \, K$ in the large field limit, where $l$ is the number of bonds separating the pair.  In other words, due to a non-vanishing transverse field, the magnetic monopoles undergo an abrupt transition into the quantum confinement regime.  Moreover, we will argue that the new spin disorder phase and the paramagnetic phase in the large field limit can be connected adiabatically by a continuous transition, and therefore the spin disorder phase at small field is a cooperative paramagnet.  In conclusion, our work serves as the \emph{simplest} model to host the spin disorder state in the 3-dimensional systems.  Finally, the effective Hamiltonian in small field will be derived for the connection of our work to the recent progress in frustrated spin systems.  

\begin{figure}[htb]
\includegraphics[width=0.48\textwidth]{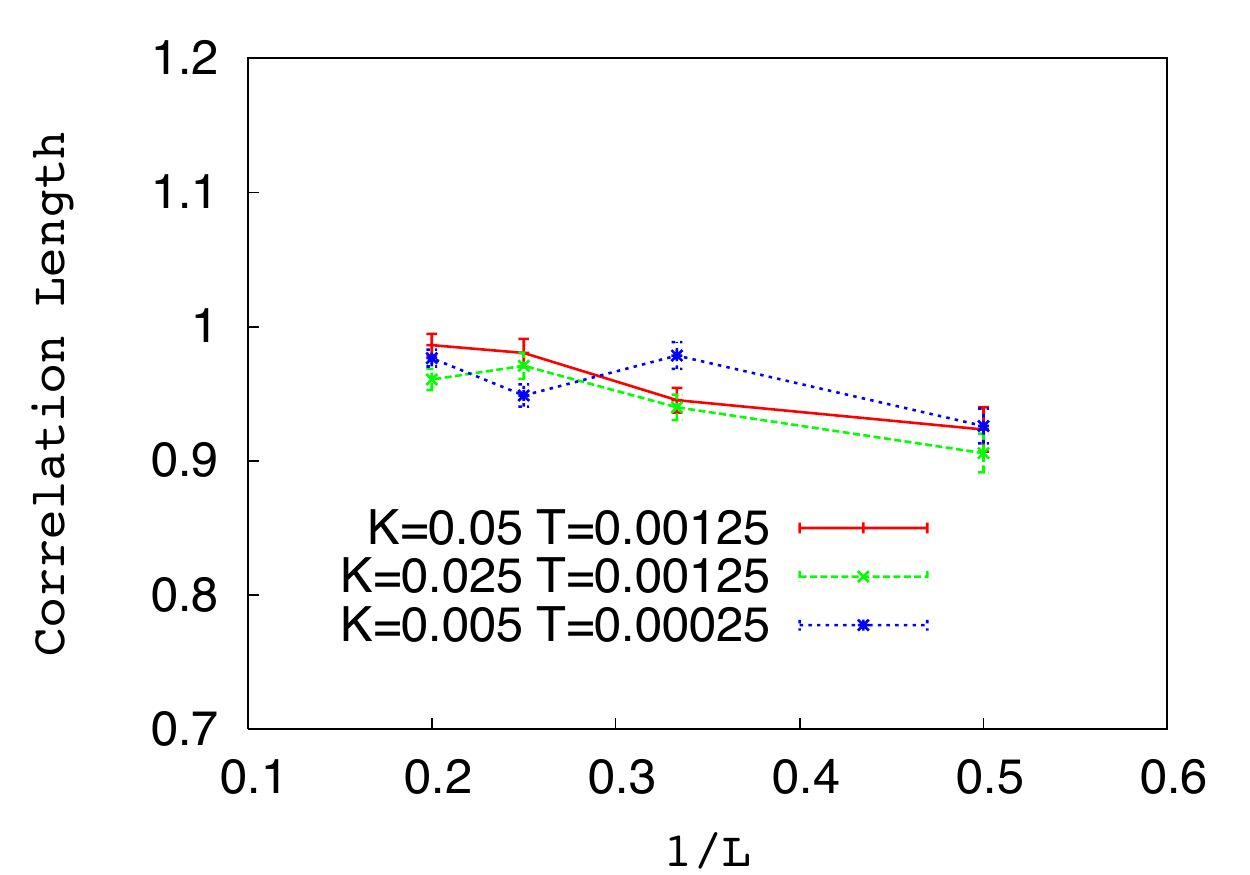}
\includegraphics[width=0.48\textwidth]{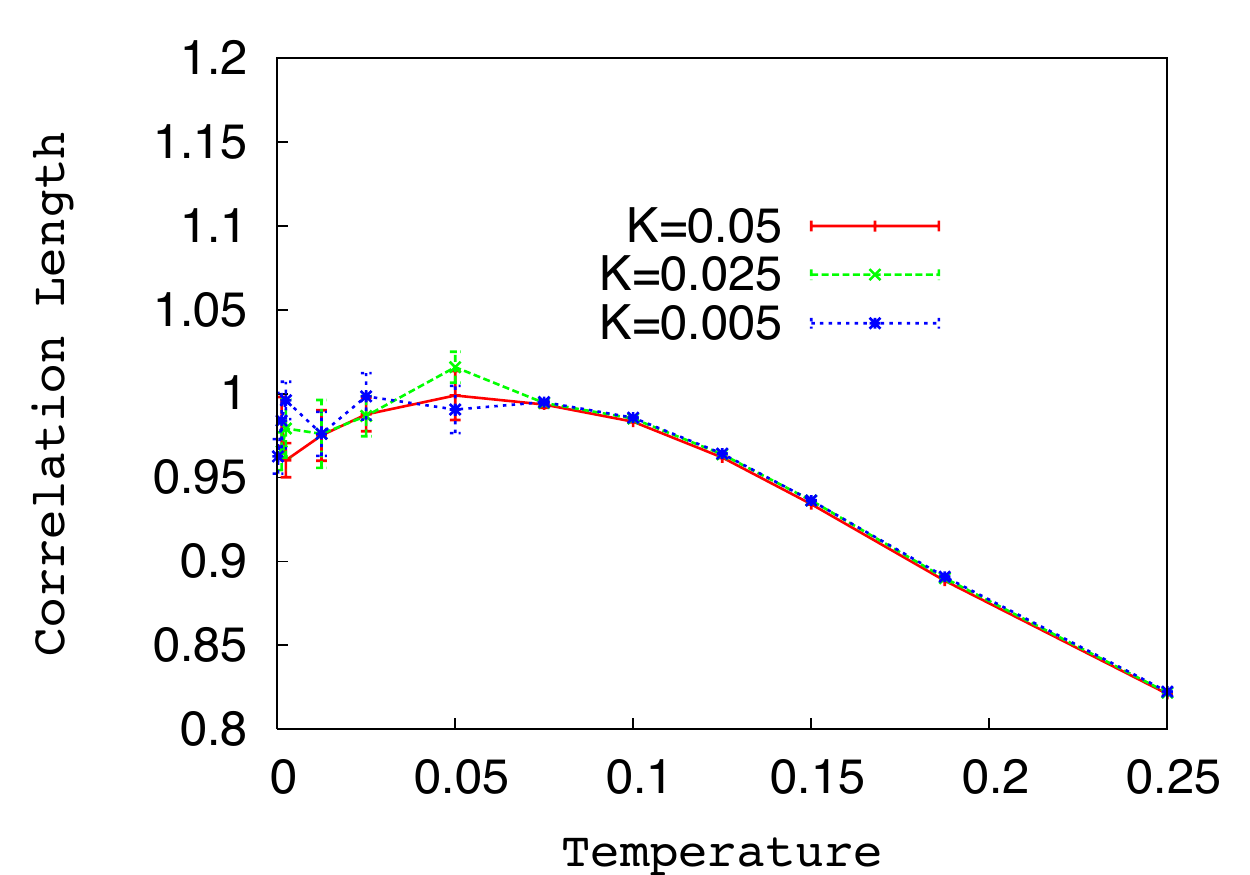}
\caption{(Color online) (a) The spin correlation length $\xi(L, T, K)$ as a function of $1/L$ for three sets of $T$ and $K$.  Both $T$ and $K$ are in units of $J$, and the correlation length is in units of the bond length.  The spin correlation is exponentially decayed and the correlation length is around one bond distance down to the temperature as low as $2.5\%$ of $K$ for $K=0.05J$ and $5\%$ of $K$ for $K=0.025J$ and $0.005J$.  The weak dimensional dependence indicates that our calculations are sufficient for the thermodynamic limit. (b) The $\xi(T,K)$ as a function of $T$ for three different $K$ values.  The correlation lengths show saturation around one bond distance at low temperature, demonstrating that the system is in the quantum disorder regime at these three field values. }\label{Fig:length}
\end{figure}

We show the disorder nature by computing the spin correlation length of $\Gamma(i)=1/N\sum_{j=1}^N\langle S^z_{i+j}S^z_j\rangle$  and the magnetization $\langle\vec{S}\rangle$ using the Trotter-Suzuki quantum Monte Carlo technique (QMC), where $\langle\rangle$ is the thermal average and $N$ is the total number of spins.  We apply the cluster algorithm in the imaginary time direction, and the heat bath algorithm in the spatial directions.  We warm up the system using $2\times10^6$ Monte Carlo steps (MCS), and perform the measurement with another $3\times10^6$ MCS.  The thermalization is checked to be good.  We take up to 64 ensemble averages using parallel computation.  We use the periodic boundary condition in $3+1$(imaginary time) directions, and compute the spin correlation length along the $z$-direction, which can be summarized by the function $\xi(n, L, T, K)$, where $n$ is the size in the imaginary time direction, $L$ is the size in the $x$ and $y$ directions, and $T$ is the temperature.  There are 16 spins in the unit cell of the pyrochlore lattice.  Our QMC results have been calibrated and agreed with the exact diagonalization results of one unit cell.  In order to obtain the correlation length in the thermodynamic limit  $\xi(T,K)=\lim_{n \rightarrow \infty,\, L\rightarrow \infty}\xi(n, L, T, K)$, scaling analysis in the imaginary time and $x$ and $y$ direction is needed. To achieve this, we compute for lattice sizes $L\times L\times z$ of $2\times2\times3$, $3\times 3\times3$, $4\times4\times2$, and $5\times5\times2$ unit cells, which contain 192, 432, 512, and 800 spins, respectively.  To reach infinite $n$ limit, we compute 4 different $n$, up to 40000 for $L=4$, and up to 30000 for $L=5$  for the smallest $T$, and apply the finite-sized scaling analysis to obtain $\xi(L,T,K)$. 

In Fig.~(\ref{Fig:length}a), we show the finite size scaling of the correlation length $\xi(L,T,K)$  as a function of $1/L$.  We compute for three different values of the transverse field, $K=0.05J$, $0.025J$, and $0.005J$.
We only plot the results of the lowest temperatures that we are able to reach as the examples.   They are as low as 2.5$\%$ of $K$ for $K=0.05J$, and 5$\%$ of $K$ for $K=0.025J$ and $0.005J$.  In many regards, these are low enough to show the ground state properties, since $K$ is the energy scale in the spectrum of low energy.  All spin correlation lengths in Fig.~(\ref{Fig:length}a) are below one bond distance from $L=2$ to $L=5$, and do not show much size dependence.  First, we observe that the spin correlation in the $z$-direction shows exponential decay with correlation length which does not change with $z$.  Second, the spin correlation shows little size dependence in the $x$ and $y$ direction.  Therefore, from the results of the short-range spin correlation and the scaling relation as shown in Fig.~(\ref{Fig:length}a), we concluded that our calculations already represent the thermodynamic limit.  $\xi(T,K)$ is computed from $\xi(L,T,K)$ by taking the mean of the values for $L=4$ and $L=5$.  We plot $\xi(T,K)$ as a function of $T$ for those field values in Fig.~(\ref{Fig:length}b).  The correlation length shows neither significant temperature dependence nor $K$ dependence, which is very similar to the kagome case~\cite{Moessner2001prb,Moessner2000prl}.  It remains very short ranged down to $T$ equal to $2.5\%$ of the field.  Therefore, the ground state remains in the spin disorder phase.  The numerical error in our calculations comes mainly from the ensemble average.  The largest error is below $\pm 3\%$ down to very low temperature.  The temperature dependence that shows the saturation of the spin correlation length implies that there is no phase transition down to zero temperature.  As a first result, the system is a spin disorder for small field, \emph{i.e.} disorder from disorder.

The next question is whether the classical monopole-anti-monopole pair is confined or deconfined in the presence of the quantum perturbation.  Let us consider the following string operator~\cite{Hermele2004prb,Castelnovo2007nature}
\begin{eqnarray}
\mathcal{O}_l = S^+_{i_1}S^-_{i_2}S^+_{i_3}S^-_{i_4}\cdot\cdot\cdot S^+_{i_l}, \label{eq:string}
\end{eqnarray}
where $i_k$ labels the sites of a connected string without intersection.  As mentioned in the introduction, $\mathcal{O}_l$ excites the a monopole and an anti-monopole at each end with the separation distance $l$ in the unit of bond length.  An example of the string excitation is given in Fig.(\ref{Fig:string}a).  They can be chosen in a random-walk way, but restricted to the condition that one tetrahedron does not contribute more than one bond to the string excitation.  The string will never intersect but can be open or closed.  A closed string corresponds to the annihilation of the monopole and anti-monopole, leaving a flux-tube-like excitation.  In the classical Ising model, a flux-tube does not cost energy.  In addition, we consider the string length $l$ to be \emph{odd} since $[M^z, \mathcal{O}_l] = \mathcal{O}_l$, where $M^z=\sum_{i=1}^NS^z_i$.  Suppose $\langle\Omega|\mathcal{O}^\dag_lM^z\mathcal{O}_l|\Omega\rangle
/\langle\Omega|\mathcal{O}^\dag_l\mathcal{O}_l|\Omega\rangle
= m_z $, where $|\Omega\rangle$ is the ground state of finite $K$, the monopole and the anti-monopole share $S^z=\frac{m_z}{2}$, {\it i.e.} fractionalization.

Now, let us consider the open string case and compute the energy of $\mathcal{O}_l$ by the following equation of motion
\begin{eqnarray}
 \!\!\!\!\!\!\!\!\!\frac{d^2\mathcal{O}_l}{dt^2} &=& -[[\mathcal{O}_l,H],H] \nonumber\\
 &=& -J^2\hat{U}_1\mathcal{O}_l-JK\hat{U}_2\mathcal{O}_l \nonumber\\&-&JK\hat{U}_3\sum_{k=1}^l(-1)^{k-1}S^+_{i_1}S^-_{i_2}\!\cdot\!\cdot\!\cdot\!S^z_{i_k}\!\cdot\!\cdot\!\cdot\!S^+_{i_l} \nonumber\\ &-& JK\sum_{k=1}^{l}(-1)^{k-1} \hat{U}_{4k}S^+_{i_1}S^-_{i_2}\!\cdot\!\cdot\!\cdot\!S^z_{i_k}\!\cdot\!\cdot\!\cdot\!S^+_{i_l} \nonumber\\ &-& \frac{1}{2}K^2\sum_{k=1}^{l}(-1)^{k-1}S^+_{i_1}S^-_{i_2}\!\cdot\!\cdot\!\cdot\!(S^+_{i_k}-S^-_{i_k})\!\cdot\!\cdot\!\cdot\!S^+_{i_l} \nonumber\\&-&2K^2\!\!\sum_{m< n}\!(\!-1\!)^{m+n}S^+_{i_1}S^-_{i_2}\!\!\cdot\!\cdot\!\cdot\!S^z_{i_m}\!\!\!\cdot\!\cdot\!\cdot\!S^z_{i_n}\!\!\!\cdot\!\cdot\!\cdot\!S^+_{i_l}, \label{eq:eom}
\end{eqnarray}
where
\begin{eqnarray}
\hat{U}_1&=& (-S^z_{i_1^*}+S^z_{i^*_2}-\cdot\!\cdot\!\cdot-S^z_{i^*_l}-(l-1))^2, \\
\hat{U}_2&=& (-iS^y_{i_1^*}+iS^y_{i^*_2}-\cdot\!\cdot\!\cdot-iS^y_{i^*_l}), \\
\hat{U}_3 &=& -S^z_{i_1^*}+S^z_{i^*_2}-\cdot\!\cdot\!\cdot-S^z_{i^*_l}-(\!l-\!1\!), \\
\hat{U}_{4k} &=&\hat{U}_3+(-1)^{k-1}S^z_{i_k^*}+h_k, \label{eq:U4}
\end{eqnarray}
where $h_k = 1$ for $k=1$ and $l$ while $h_k=2$ for others, and $i_k^*$ label the nearest neighbor spins of the $i_k$ site.  In our system, one site has 6 neighbors.  So far, the equation of motion as given through Eq.(\ref{eq:eom}) to Eq.(\ref{eq:U4}) is exact.  We compute the energy in the $K \ll J$ limit by applying mean field theory.  Take the following expectation values:
\begin{eqnarray}
&&\langle\hat{U}_1\rangle_{\mathcal{O}_l} = \langle(-S^z_{i'_1}-S^z_{i'_l})^2\rangle_{\mathcal{O}_l} \equiv \Delta_0,\nonumber \\ &&\langle\hat{U}_2\rangle_{\mathcal{O}_l} \simeq 0, \nonumber \\
&&\langle(-1)^{k-1}(\hat{U}_3+\hat{U}_{4k})S^z_{i_k}\rangle_{\mathcal{O}_l}\! \nonumber \\&&=\left\{\begin{array}{l}
\langle-S^z_{i'_1}\!-\!S^z_{i'_l}\!+\!\frac{1}{2}S^z_{i^*_1}\!+\!\frac{1}{2}\rangle_{\mathcal{O}_l}\ for \ k=1\\
\langle-S^z_{i'_1}\!-\!S^z_{i'_l}\!+\!\frac{1}{2}S^z_{i^*_l}\!+\!\frac{1}{2}\rangle_{\mathcal{O}_l}\ for\ k=l \\ \langle-S^z_{i'_1}\!\!-\!S^z_{i'_l}\!+\!\frac{1}{2}(\!-1)^{k\!-\!1}S^z_{i^*_k}\!\!+\!1\rangle_{\mathcal{O}_l}\\ for\ k=2,3,..,l-1,\end{array}\right\}\equiv \rho_k, \nonumber \\&&\langle S^z_{i_m}S^z_{i_n}\rangle_{\mathcal{O}_l}=\frac{1}{4}(\!-1\!)^{m+n}, \label{eq:meanfield}
\end{eqnarray}
where $\langle\rangle_{\mathcal{O}_l}$ denotes the expectation value of one string state, $i'_1$ labels the sites of the first tetrahedron without $i_1$, and likewise for $i'_l$. Eq.~(\ref{eq:eom}) becomes
\begin{eqnarray}
\!\!\!\!\frac{d^2\mathcal{O}_l}{dt^2} &+& \left(J^2\!\Delta_0\!+\!\frac{1}{2}lK^2\right)\mathcal{O}_l=-JK\sum_{k=1}^l\rho_k\mathcal{O}'_{k,l-1}\nonumber\\&-&\frac{1}{2}K^2\sum_{m\neq n}\!\!\mathcal{O}''_{mn,l-2} \label{eq:eommf}
\end{eqnarray}
where $\mathcal{O}'_{k,l-1}$ denotes the string operator of bond length $l-1$ without $S^{\pm}_{i_k}$ in the $\mathcal{O}_l$, and $\mathcal{O}''_{mn,l-2}$ are the ones of bond length $l-2$ without $S^{\pm}_{i_m}$ and $S^{\pm}_{i_n}$ in the $\mathcal{O}_l$.

Eq.~(\ref{eq:eommf}) is nothing but the simple harmonic equation for $\mathcal{O}_l$ with the eigenvalue $\omega=\sqrt{J^2\!\Delta_0\!+\!\frac{l}{2}K^2}$ coupling to the strings of length $l-1$ and $l-2$.  It can be solved by computing the equations of motion (EOM) for $\mathcal{O'}_{k,l-1}$ and $\mathcal{O}''_{mn,l-2}$, and obtain the set of EOM for the strings from $l$ to the unit lengths.  Solving the set of EOM bottom up, the right hand side of Eq.~(\ref{eq:eommf}) becomes a large number of the quantum fluctuations of the string excitations of all lengths from $l-1$ to unity.  For example, in Eq.~(\ref{eq:eommf}) there are $l$ $\mathcal{O}'_{k,l-1}$ and $l(l-1)/2$ $\mathcal{O}''_{mn,l-2}$ fluctuations, and they have the equations of motion of $l-1$ string fluctuations of $l-2$ length, and $(l-1)(l-2)/2$ string fluctuations of $l-3$ length, and so on.   One can safely assume that  their phases are random so that the inhomogeneous part in Eq.~(\ref{eq:eommf}) is not crucial.   Then, the string excitation $\mathcal{O}_l$ is the quasi-eigenstate with the energy $\omega$ for small $K$.  The procedure given above is the generalized random phase approximation for the string-like excitation $\mathcal{O}_l$.

The monopole-anti-monopole pair is massive, because a single spin excitation takes the system away from the ground state manifold, which corresponds to the string excitation of unit length.  The mass gap of the monopole--anti-monopole pair is $J\sqrt{\Delta_0}=J$ for $K=0$.  Because there is macroscopic degeneracy in the ground state manifold, and also because the closed flux tube connects different ground states, the system is deconfined at the classical level.  Our result of $\omega$ is consistent with this scenario.  For finite $K$, $\omega$ is $l$-dependent.  It can be understood as the mass gap from the $J$ term and the binding energy from the $K$ term.  When $K$ is small, the binding energy is proportional to $\l\frac{K^2}{J}$.  As the binding energy is proportional to the separation of distance, the monopole-anti-pole pair is in a confined phase in the presence of the quantum perturbation.  It is because the operator $\mathcal{O}_l$ actually is a projection operator for spin-$1/2$.  As the ground state is spin disorder, namely $<m_z>=0$ at every site, the operator creates a sequence of ordering sites which is nothing but the Dirac string of which the energy is proportional to its length.

We note that the slope of the confining potential is small when $K\ll J$.  It is nearly deconfined.  Although the open string excitation is an energetically costly quasi-eigenstate, the flux-tube excitation of the closed string is less expensive.  In the pyrochlore lattice, the shortest closed string is the pyrochlore hexagon the energy of which is $\sqrt{3}K$.  In general, the energy of the closed string is $\sqrt{\frac{l}{2}}K$.  Furthermore, we remark that the mechanism of the quantum confinement by the transverse field is non-trivial.  As it is in the one-dimensional ferromagnetic Ising model, spinons are deconfined in the classical case and remain deconfined when the transverse field is turned on.  It is the longitudinal field that makes it confined.

For $K\gg J$, the mean field values of Eq.~(\ref{eq:meanfield}) are no longer valid.  The $K^2$ terms dominate in the EOM, and $\mathcal{O}_l$ is far from being the eigenstate.  However, we can still compute the energy by $\langle\Phi|\mathcal{O}_l^\dag H \mathcal{O}_l|\Phi\rangle/\langle\Phi|\mathcal{O}_l^\dag\mathcal{O}_l|\Phi\rangle$, where $|\Phi\rangle$ is the ground state in the large $K$ limit.  It is not hard to obtain the leading term of the energy $\frac{1}{2}lK$.  It is not surprising that the paramagnetic phase is a confined phase.  Therefore, this mode is confined for any finite $K$, which implies that the paramagnetic phase in small $K$ undergoes a continuous transition to the one in the large $K$ limit.  In other words, the phase in the small $K$ is a cooperative paramagnet.

Finally, we compute the effective Hamiltonian for small $K$ using the Rayleigh-Schr\"odinger degenerate perturbation theory, which is the following:
\begin{eqnarray}
\!\!\!\!\!\!\!\!\!\!H_{eff} \!= \!J\,\!\!\!\sum_{\langle ij\rangle}\!\!S^z_iS^z_j \!+\! J_{ring}\,\!\sum_{\hexagon_j}\left(S^-_{j_6}S^+_{j_5}S^-_{j_4}S^+_{j_3}S^-_{j_2}S^+_{j_1} \!\!+\! h.c.\!\right) \label{eq:effectiveH}
\end{eqnarray}
where $J_{ring}=\frac{63K^6}{16J^5}$, $j_k$ labels the $k^{th}$ site of the hexagon $j$, and the second summation is over all hexagons in the pyrochlore lattice.  An example of a hexagon is given in Fig.~(\ref{Fig:string}a).  Eq.~(\ref{eq:effectiveH}) is the classical Ising term plus a ring exchange term, which was previously studied by Hermele et al.~\cite{Hermele2004prb}.  They found the deconfined phase of the emergent pyrochlore photon in the large $J_{ring}$ limit.  The present work fills the blank in the small $J_{ring}$ region.  We note that the large $J_{ring}$ limit is different from the large $K$ limit.  Nevertheless, although Eq.~(\ref{eq:effectiveH}) describes the physics of small-$K$, the open monopole pair is not in its spectrum, since it is massive in the order of $J$.  For a close string, the $J_{ring}$ term is proportional to its rotational energy moving around the pyrochlore hexagon.

In summary, our work introduces a new cooperative paramagnet in the three-dimensional pyrochlore system.  The antiferromagnetic quantum Ising model in the pyrochlore lattice is a spin disorder ground state, and the monopole-anti-monopole pair is confined.  The deconfinement only occurs at the $K=0$ point.  Unlike the perturbation of the classical dipolar interaction, the quantum perturbation by the transverse field introduces the \emph{quantum confinement}.  The disorder phase adiabatically connects to the paramagnetic phase in the large field limit.  Some interesting research directions include the finite temperature properties, the computation of experimentally measurable quantities, and search for new quantum spin liquid in other 3-dimensional systems.

CHC is grateful for the discussion with Peter Fulde.  Special gratitude is heartily given to Naoto Nagaosa for leading him to this field.  CHC and CNL are supported by NSC 97-2112-M-002-027-MY3 of Taiwan.

\end{document}